\documentclass{he_symp}
\usepackage{psfig}
\newcommand{\gapr}{\raisebox{-.6ex}{\mbox{
$\stackrel{>}{\mbox{\scriptsize$\sim$}}\:$}}}
\newcommand{\lapr}{\raisebox{-.6ex}{\mbox{
$\stackrel{<}{\mbox{\scriptsize$\sim$}}\:$}}}
\def\ros{{\sl ROSAT}}
\def\asc{{\sl ASCA}}
\def\cha{{\sl Chandra}}
\begin{document}
\title{
Thermal Radiation from Neutron Stars: Chandra Results
}
\author{G.G. Pavlov\inst{1}, V.E. Zavlin\inst{2} \and D. Sanwal\inst{1}
}
\institute{The Pennsylvania State University, 525 Davey Lab.,
University Park, PA 16802, USA
\and Max--Planck--Institut f\"ur extraterrestrische Physik,
 Giessenbachstra{\ss}e, 85748 Garching, Germany}
\maketitle

\begin{abstract}
The outstanding capabilities of the {\sl Chandra} X-ray observatory
have greatly increased our potential to observe and analyze
thermal radiation from the surfaces of neutron stars (NSs).
Such observations allow one to measure the surface temperatures
and confront them with the predictions of the NS cooling models.
Detection of gravitationally redshifted spectral lines can yield
the NS mass-to-radius ratio.
In rare cases when the distance is known, one can measure
the NS radius, which is particularly important
to constrain the equation of state of the superdense matter
in the NS interiors. Finally, one can infer
the chemical composition of
the NS surface layers, which provides information about
formation of NSs and their interaction with the
environment.
We will overview the recent {\sl Chandra} results 
on the thermal radiation from various types of NSs
--active pulsars, young radio-quiet neutron stars in supernova remnants,
old radio-silent ``dim'' neutron stars-- and  discuss their
implications.
\end{abstract}
\section{Introduction}
Observational study of 
thermal emission from isolated (non-accreting) neutron stars (NSs)
became possible after the launch of
{\sl Einstein} (1979) and {\sl EXOSAT} (1983),
the first X-ray observatories equipped with X-ray telescopes.
Contrary to nonthermal radiation of NSs, generated in
the pulsar magnetospheres and observed
from radio to $\gamma$-rays, thermal emission
originates immediately at the surface, with the bulk of the
energy flux 
in the soft X-ray band.
Therefore, comparing the spectra and the light curves
with the models
for NS thermal radiation (see the contribution
by Zavlin \& Pavlov in
these Proceedings; ZP02 hereafter),
one can infer the NS surface
temperature, magnetic field, gravitational acceleration,
and chemical composition, as well as the NS mass and radius.
The analysis of 
such observations
allows one to trace thermal evolution of NSs (Tsuruta 1998)
and constrain 
the properties of the superdense matter in the NS interiors 
(D.G.\ Yakovlev et al., these Proceedings).

Observational manifestations of NSs
are very diverse, 
and only some of these exotic objects are
suitable for observing the NS
thermal emission.
Most of the {\em detected} NSs are radio
pulsars.
Very young, active pulsars (with ages $\tau\lapr 10^3$ yr ---
e.g., the Crab pulsar)
are quite hot, with expected surface temperatures $\sim 1$--2 MK,
but their nonthermal radiation is so bright
that the thermal radiation is hardly observable.
The nonthermal component in X-ray emission of middle-aged
pulsars ($\tau\sim 10^4$--$10^6$ yr --- e.g., 
B0656+14, B1055--52) is much fainter, and
their thermal radiation, 
with temperatures 0.3--1 MK, can dominate at soft X-ray and UV energies.
The surfaces of old pulsars ($\tau\gapr 10^6$ yr --- e.\ g., 
B0950+08, B1929+10, J0437--4715) are
too cold to be seen in X-rays, but their polar caps are expected to be
heated up to a few million kelvins,
hot enough to emit observable thermal X-ray
radiation.

In addition to active pulsars, a number of radio-quiet
isolated NSs emitting thermal-like X-rays have been detected,
with typical temperatures $\sim 0.5$--5 MK.
They are usually subdivided in four classes: Anomalous X-ray
Pulsars (AXPs; see 
S.\ Mereghetti et al., these
Proceedings), Soft Gamma-ray Repeaters (SGRs; Kouveliotou 1995),
``dim'' or ``truly isolated'' radio-silent NSs (i.e., not associated with
supernova remnants [SNRs]; Treves et al.\ 2000), and compact
central sources (CCOs) in SNRs (Pavlov et al.\ 2002a) which have been
identified with neither active pulsars nor AXPs/SGRs.
Their observational manifestations
(particularly, multiwavelength spectra) are quite different from those of
``active'' pulsars, and their properties 
have not 
been investigated as extensively,  
but the presense of the thermal
component in their radiation provides a clue to understand the nature
of these objects.

First thermally emitting isolated NSs
were detected with
{\sl Einstein}
(e.g., Fahlman \& Gregory 1981; Cheng \& Helfand 1983;
Helfand \& Becker 1984; Seward, Charles, \& Smale 1986;
C\'ordova et al.\ 1989) and investigated with {\sl EXOSAT}
(Brinkmann \& \"Ogelman 1987; Kellett et al.\ 1987).
In 1990's, \ros\ and \asc\
detected X-rays from more than 30 radio pulsars,
including a few thermal emitters (\"Ogelman 1995), and discovered 
several radio-silent NSs
 (see Becker \& Pavlov 2002 for a review).
Thermal emission from a few 
isolated NSs
was also detected in the optical-UV range
with the {\sl Hubble Space Telescope} (e.g., Pavlov, Stringfellow \&
C\'ordova 1996; 
 Walter \& Matthews 1997).

The new era in observing X-ray emission from NSs has
started
with the launch of two currently operating X-ray observatories
of outstanding capabilities, {\sl Chandra} and {\sl XMM}-Newton.
In this review we present recent results on 
thermal X-ray emission from isolated NSs of
various types observed with \cha.
A more general overview of \cha\ observations of NSs is presented
in this volume by
M.C.\ Weisskopf, while the {\sl XMM}-Newton results on NSs are 
reviewed by W.\ Becker.

\begin{figure}[ht]
\centerline{\psfig{file=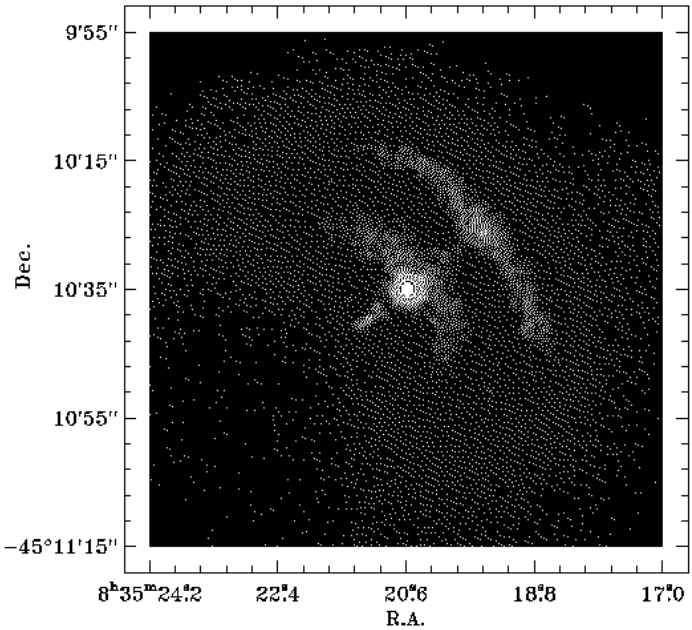,height=10cm,clip=} }
\caption{
\cha\ HRC-I image of the Vela pulsar and surrounding nebula.
}
\label{velaimg}
\end{figure}

\begin{figure}
\centerline{\psfig{file=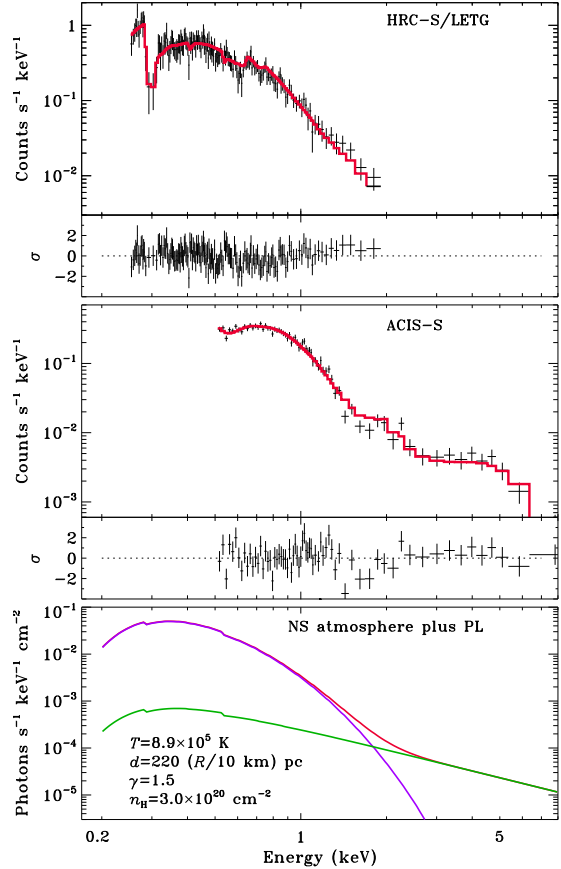,height=12cm,clip=} }
\caption{
Two-component (magnetic hydrogen
NS atmosphere plus power law) model fit to the combined
HRC-S/LETG and ACIS count rate spectra.
The bottom panel
shows the contributions from the thermal and nonthermal
components (lilac and green lines, respectively).
}
\label{velaspec1}
\end{figure}

\begin{figure}
\centerline{\psfig{file=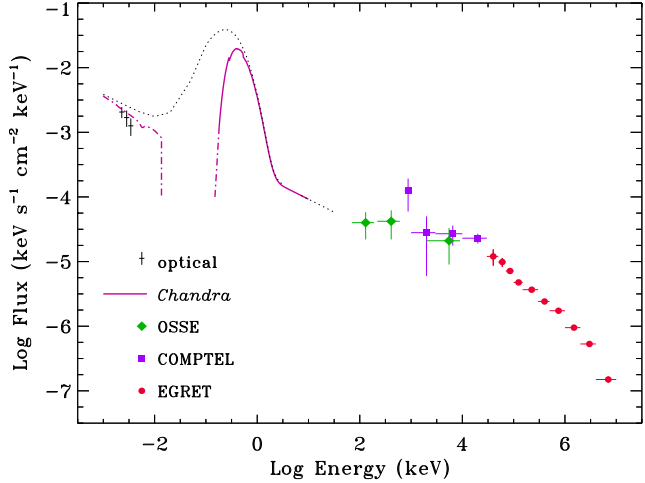,height=7cm,clip=} }
\caption{
Multiwavelength energy spectrum of the Vela pulsar.  The solid line
shows the NS atmosphere plus PL fit to the observed LETG and ACIS spectra.
The dotted line corresponds to same model spectrum corrected for the
interstellar absorption and extrapolated to lower and higher energies.
The dash-dot lines show the extrapolated optical and EUV absorbed spectra
(see Pavlov et al.~2001b for 
details and references).
}
\label{velaspec2}
\end{figure}

\begin{figure}
\centerline{\psfig{file=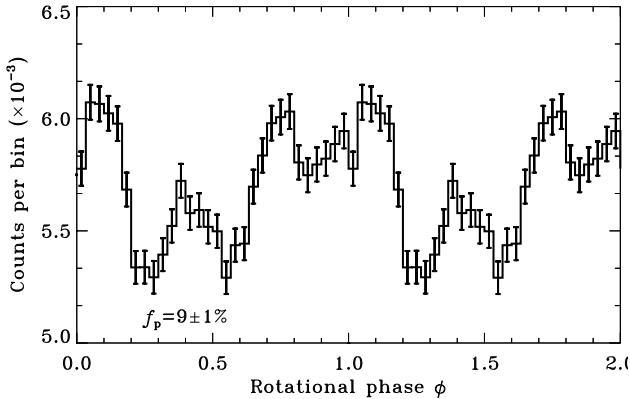,height=6cm,clip=} }
\caption{Light curve of the Vela pulsar from two HRC-I observations
of January and February 2000.
The zero phase corresponds to the radio peak.
}
\label{velalc}
\end{figure}

\section{Thermal emission from radio pulsars}
X-ray emission of rotation-powered pulsars generally consists
of two components --- thermal and nonthermal. The nonthermal component,
generated by relativistic particles in the NS
magnetosphere,
is, as a rule, strongly pulsed
and characterized by a power-law (PL) spectrum.
Studying the magnetospheric component
is important to understand
the mechanisms of the multiwavelength (optical through gamma-rays) pulsar
radiation. The thermal component, emitted from the NS surface,
 can be observed in X-rays if it is not
buried under the nonthermal component, as in very young pulsars,
and if the temperature is not too low, as in old pulsars.
Weak pulsations of the thermal emission can be caused by temperature
nonuniformities along the NS surface
and/or anisotropy of the radiation intensities in a strong magnetic field
(see ZP02). Stronger pulsations of thermal
radiation can be observed in old pulsars which emit thermal X-ray radiation
from small, hot
polar caps around the magnetic poles.
In this Section we present the results of our analysis
of \cha\ data on pulsars of different ages:
the young Vela pulsar ($\tau_c\equiv P/2\dot{P}=11$ kyr), 
the middle-aged pulsars B0656+14
($\tau_c=110$ kyr) and 
B1055--52 ($\tau_c=500$ kyr),
and the very old millisecond pulsar J0437--4715 ($\tau_c=4.9$ Gyr). 
\subsection{The Vela pulsar}
First \cha\ observations of this
famous 89-ms radio, optical and $\gamma$-ray pulsar in the Vela
SNR (the distance $d\approx 300$ pc --- Caraveo et al.\ 2001)
were carried out with the imaging array of the High Resolution Camera (HRC-I)
in January and February 2000. These observations revealed 
a very complicated structure of the PWN around the pulsar 
(see Fig.~\ref{velaimg})
studied by Helfand, Halpern \& Gotthelf (2001) and Pavlov et al.~(2000, 2001a).
The superb spatial resolution
of \cha\ made it possible, for the first time, 
to resolve the pulsar from its surroundings
and investigate the pulsar's radiation.

The spectral data on the Vela pulsar were obtained in two observations
(Pavlov et al.~2001b).
A 25 ks HRC-S
observation with Low Energy Transmission Grating (LETG) of January 2000 
provided the high-resolution
pulsar spectrum 
in the energy range $E=0.25$--2.0~keV.
The analysis of the
LETG spectrum ($\approx 7,000$ source counts)
did not reveal statistically significant spectral lines.
The spectrum is best fitted with thermal models (blackbody, 
hydrogen NS atmosphere), but the observed spectrum somewhat exceeds
the model spectra at 
$E\gapr 1.5$~keV
(see upper panel in Fig.~\ref{velaspec1}),
indicating the presence of a second component with a harder spectrum.
The spectrum at higher photon energies
was provided by a 37 ks 
observation (October 1999) with the Advanced CCD Imaging Spectrometer (ACIS), 
taken in the Continuous Clocking (CC) mode that allows timing at the expense of
imaging capability in one dimension.
The 1-D count distributions of the pulsar and the nebula
show that the pulsar is much brighter than the PWN background
at $E\lapr 1.5$ keV, and it is discernible up to 8~keV, where the charged
particle background takes over (see Fig.\ 1 in Pavlov et al.\ 2001b).
The pulsar spectrum in the 0.5--8 keV band
($\approx 9,000$ source counts) does not fit 
one-component models,
but it fits fairly well 
two-component models
(e.g., a PL plus a thermal component --- see the middle panel in Fig.\ 2).

The analysis of the combined (LETG plus ACIS) spectra proves that
the pulsar's soft X-ray emission, at $E\lapr 1.8$~keV, is indeed dominated
by a thermal component (see bottom panel in Fig.~\ref{velaspec1}).
The parameters of this component are drastically different
for the blackbody fit ($T^\infty_{\rm bb}=1.4$--1.5~MK, $R^\infty_{\rm bb}=
[2.3$--$2.9]\,d_{300}$ km) and the NS hydrogen atmosphere fit
($T_{\rm eff}^\infty =0.65$--0.71~MK, $R^\infty =
[17$--$20] d_{300}$ km).
Such strong difference is explained by the 
fact that
the spectrum of a hydrogen atmosphere
decreases with increasing photon energy slower than the ideal Wien spectrum because
the radiation at higher energies comes from hotter layers.
As a result, the blackbody fit
to a hydrogen atmosphere spectrum yields a temperature higher
than the true effective temperature and an area smaller
than the true emitting area (see ZP02
for more details).

The small blackbody radius 
might hint that
the observed radiation is emitted
from hot spots (e.g., polar caps heated by relativistic particles
produced in the pulsar magnetosphere), whereas the radius obtained
in the hydrogen atmosphere fit implies radiation emitted from the entire NS
surface. 
Rigorously speaking, the surface (atmosphere) of a NS
is not a black body, but a blackbody
spectrum could mimic the spectrum of a heavy-element (e.g., iron)
atmosphere if it is observed with low spectral resolution.
However, the lack of significant spectral
lines in the LETG spectrum indicates that there are no heavy elements,
neither He nor metals, in the radiating atmospheric layers.
On the other hand, even strongly
magnetized hydrogen does not have spectral features in the investigated
energy band (all the hydrogen features are below 0.17~keV for
the field of $B=3\times 10^{12}$ G, adopted for the Vela pulsar).
Therefore, the hydrogen atmosphere interpretation looks more plausible.
With this interpretation,
the effective temperature of the Vela pulsar is well below the predictions
of the ``standard'' models of the NS cooling (e.g., Tsuruta 1998).

An important result of this analysis is the detection of the
hard spectral component which dominates at $E\gapr 1.8$~keV.
The slope (photon index) of this PL component is $\gamma=1.2$--1.8
if the NS hydrogen atmosphere model is adopted for the soft component.
With this interpretation,
the overall X-ray spectrum of the the Vela pulsar is
quite similar to those of the three middle-aged pulsars,
Geminga, B1055--52 and B0656+14 (see below), where the
thermal component was detected, although the ratio of the nonthermal X-ray
luminosity to the spin-down energy loss rate,
$L_{\rm x}/\dot{E}\simeq 
3\times 10^{-6}$ in the 0.2--8.0~keV range, is a
factor of $10^3$ smaller than observed for typical X-ray detected
pulsars (Becker \& Tr\"umper 1997).  If the $L_{\rm x}/\dot{E}$ ratio
were as high as for other pulsars, the thermal radiation would be undetectable
in the phase-integrated spectrum.

It is illuminating to compare the nonthermal X-ray spectrum of
the pulsar with those in other energy ranges. Figure~\ref{velaspec2} shows the
spectral flux of the pulsar from optical to gamma-rays.
Remarkably, the extrapolation of the PL spectrum obtained in the
atmosphere-plus-PL fit matches fairly well with the optical and hard-X-ray
(soft-$\gamma$-ray) time-averaged fluxes.  Moreover, under
the hypothesis that the PL spectrum has the same slope
in the optical through X-ray range, we can constrain
the photon index very tightly: $\gamma = 1.35$--1.45.

The multicomponent structure of the pulsar radiation is also seen from
its pulse profile.
The 
HRC-I light curve is shown in
Figure~\ref{velalc} (see also Helfand et al.\ 2001; Sanwal et al.\ 2002a).
It has at least three peaks --- quite unusual for a pulsar, but not so
surprising for the Vela pulsar which shows up to four
peaks in the hard X-ray
bands observed with {\sl RXTE} (Harding et al.~2002).

So far, only the analysis of phase-integrated spectra and
energy-integrated light curves has been possible.
Phase-resolved spectral analysis and energy-resolved
timing 
would be of crucial
importance to elucidate the complicated emission mechanism(s) of 
both the nonthermal and thermal components.
In particular, the analysis of the phase-energy dependence of the thermal
emission, based on the NS atmosphere models
can make it possible to determine the surface distributions of the
temperature and the magnetic field.

\subsection{Middle-aged pulsars B0656+14 and B1055--52}
Among about 40 rotation-powered pulsars detected with
the previous and currently operating X-ray missions,
middle-aged pulsars ($\tau\sim 0.1$--1 Myr)
are of special 
interest because their thermal emission is expected to dominate
in soft X-rays.
The best studied objects of this class are the
``Three Musketeers''\footnote{As dubbed by J.~Tr\"umper}:
PSR B0656+14, Geminga, and PSR B1055--52.

\subsubsection{PSR B0656+14}
PSR B0656+14 ($\tau_c=
110$ kyr) is 
the brightest of the Three Musketeers.
Observations with {\sl Einstein}, \ros, and \asc\
have shown that its X-ray spectrum
consists
of at least two components. The thermal 
component
($E\lapr 1$ keV) resembles a soft blackbody spectrum with an
apparent temperature of about 0.9 MK.
Because of the narrow energy band of \ros\
and poor angular resolution of \asc,
the shape of the hard component (at $E\gapr 1$~keV)
could hardly be determined from those data. For
instance, the \asc\ spectra above 1~keV 
can be fitted equally well with
a hard blackbody with temperature of about 2.5--4.0 MK
or a PL with  photon index $\gamma = 1.3$--3.0.
IR-optical-UV observations of the pulsar
(Pavlov et al.\ 1997; Koptsevich et al.\ 2001) revealed a nonthermal
PL component with $\gamma = 1.4$--1.5,
presumably emitted from the pulsar magnetosphere.

\begin{figure}
\centerline{\psfig{file=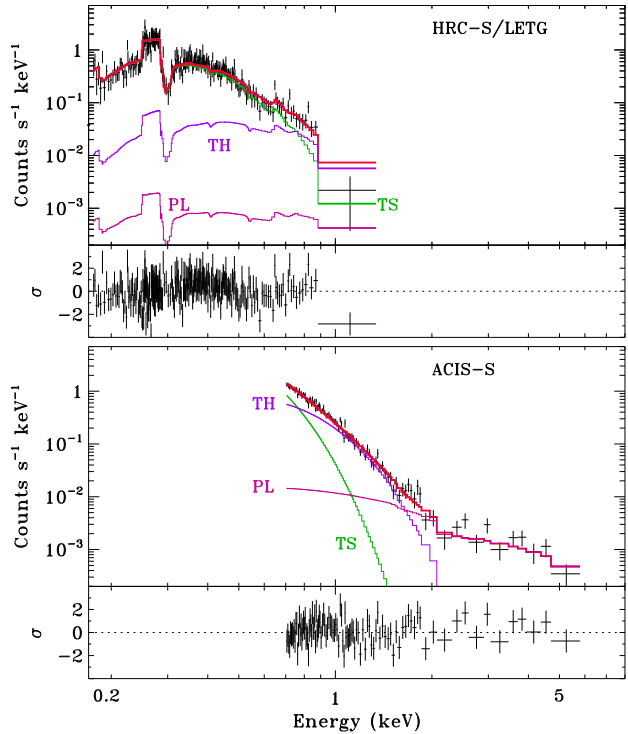,height=10cm,clip=} }
\caption{Three-component, TS+TH+PL,
fit to the combined
LETG and ACIS-S spectra of
PSR B0656+14. 
The red curves show the 
model countrate spectra
for the two instruments.
}
\label{0656spec1}
\end{figure}

\begin{figure}
\centerline{\psfig{file=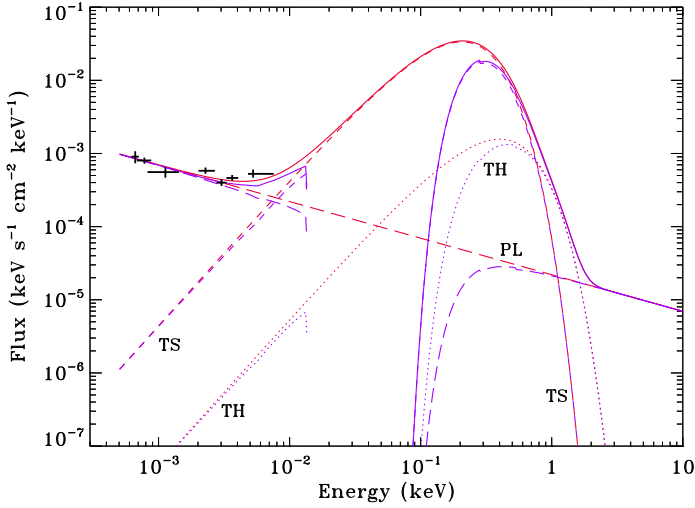,height=8cm,clip=} }
\caption{Spectral flux for the TS+TH+PL model 
for PSR B0656+14 extrapolated to the optical range.
Blue and red curves show
the absorbed and unabsorbed 
spectra,
respectively. 
Crosses are the IR-optical-UV fluxes 
(Koptsevich et al.\ 2001).
}
\label{0656spec2}
\end{figure}

\begin{figure}
\centerline{\psfig{file=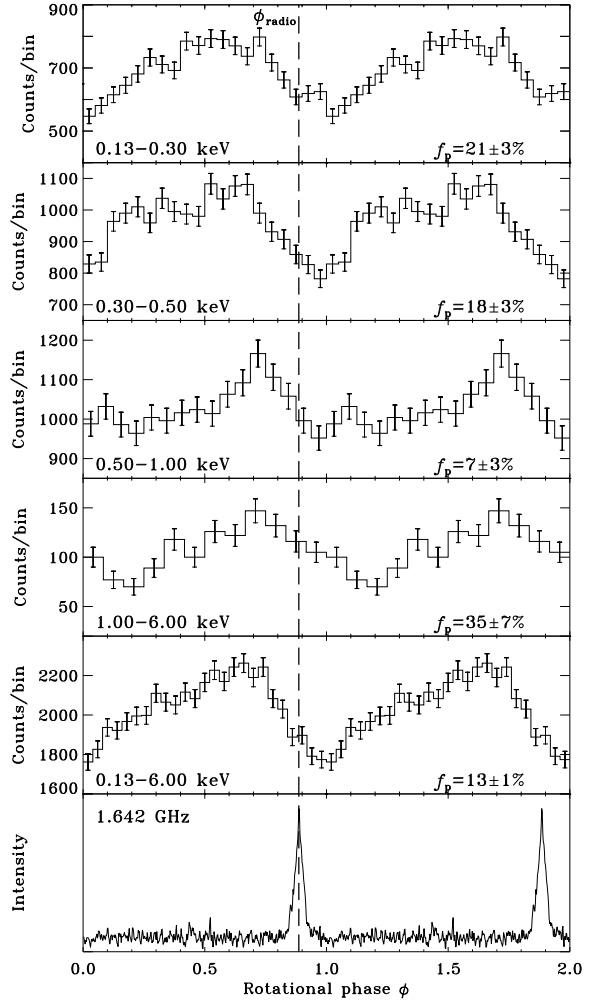,height=14cm,clip=} }
\caption{X-ray pulse profiles
for PSR B0656+14
in different energy bands (four upper panels).
The radio pulse profile at 1.642 GHz (Gould \& Lyne 1998) demonstrates
the phase
difference between the X-ray and radio peaks.
The X-ray data are from the \cha\ ACIS observation
in the CC
mode (December 2001).
}
\label{0656lc}
\end{figure}

\begin{figure}[ht]
\centerline{\psfig{file=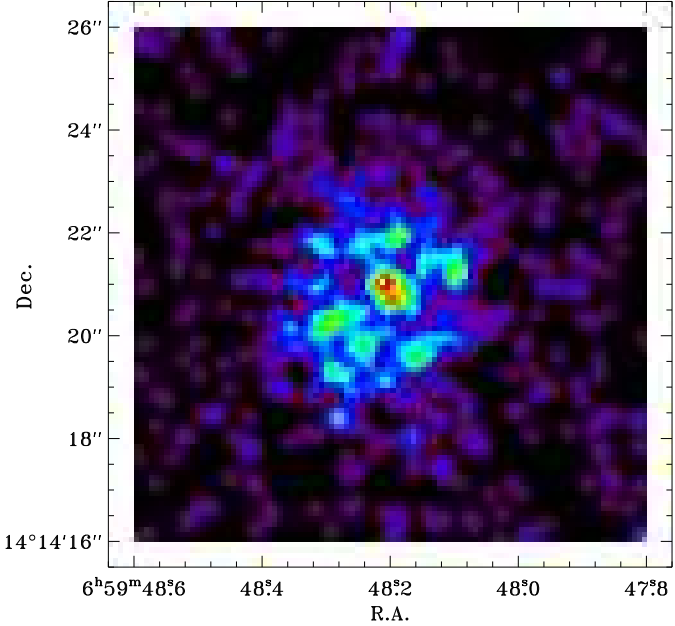,height=10cm,clip=} }
\caption{
Image of a $10''\times 10''$ area around PSR B0656+14
from the 5~ks ACIS observation in the Timed Exposure mode
(December 2001). 
}
\label{0656img}
\end{figure}

\begin{figure}
\centerline{\psfig{file=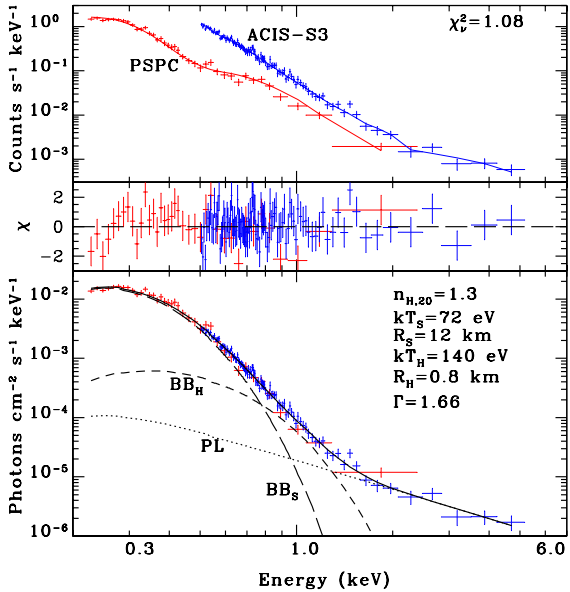,height=8cm,clip=} }
\caption{Three-component model fit 
to the \cha\ ACIS and \ros\ PSPC
spectra of PSR B1055--52.
}
\label{1055xraysp}
\end{figure}

\begin{figure}
\centerline{\psfig{file=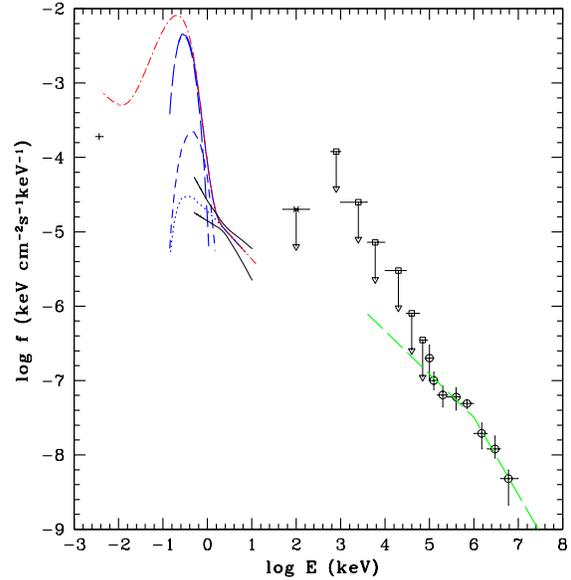,height=8cm,clip=} }
\caption{Spectral flux from PSR 1055--52 for the TS+TH+PL model
extrapolated to the optical and $\gamma$-ray ranges,
together with the data from {\sl HST}, {\sl OSSE}, {\sl COMPTEL} and
{\sl EGRET}.
The blue and the red curves give the absorbed and unabsorbed spectra,
respectively.
}
\label{1055spec}
\end{figure}

\begin{figure}
\centerline{\psfig{file=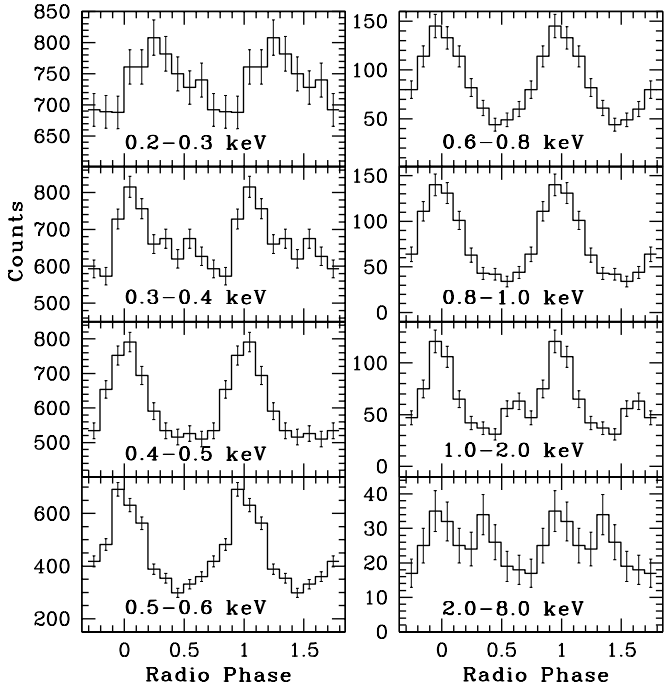,height=10cm,clip=} }
\caption{
Energy-resolved light curves of PSR B1055--52 plotted
vs.\ radio phase.
}
\label{1055lc}
\end{figure}

\begin{figure}[!hb]
\centerline{\psfig{file=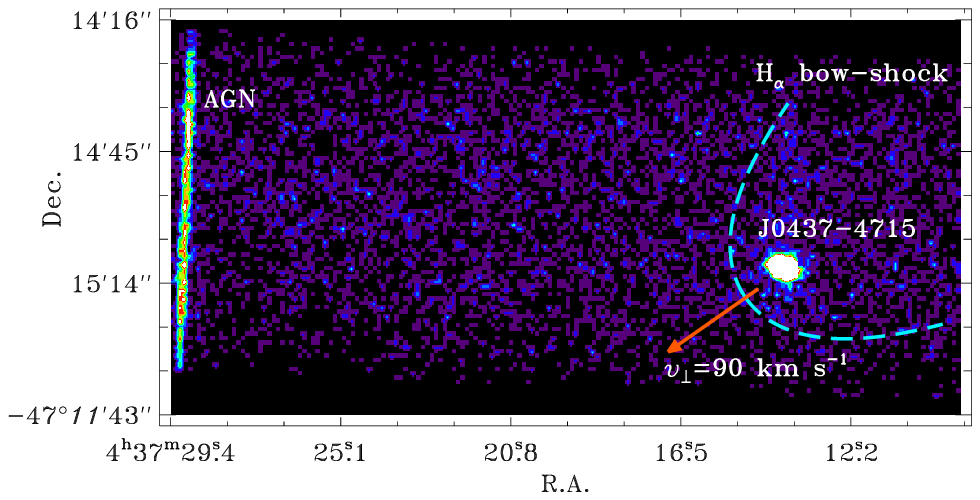,height=6cm,clip=} }
\caption{
ACIS image of PSR 0437--4715. The bright strip is the trailed image
of a nearby AGN. The dashed curve marks the leading edge
of the H$_\alpha$ bow-shock. The arrow shows the direction
of the pulsar's proper motion.
}
\label{0437img}
\end{figure}

To search for spectral features in the X-ray spectrum of PSR B0656+14
and investigate the shape
of its high-energy tail, the pulsar
was observed with HRC-S/LETG in November 1999 (38 ks
exposure) and ACIS-S (in CC
mode) in December 2001
(25 ks exposure). The analysis of these data showed 
no significant 
features in the source spectrum.
To obtain a good fit to the combined X-ray spectrum in the 0.2--6 keV band,
a three-component model is required.
The high-energy tail ($E\gapr 2$ keV) fits best with a PL spectrum
with a slope close to that of the optical spectrum.
To constrain it more tightly, we assumed that the slope does not change from
optical to X-rays, included the optical data in the fit,
and obtained $\gamma = 1.5$. The low-energy part of the spectrum
($E\lapr 0.7$ keV) fits well with a blackbody spectrum with temperature
$T_{\rm s}^\infty = 0.85$ MK and equivalent radius of emitting sphere
$R_{\rm s}^\infty=12.0\,(d/500\, {\rm pc})$ km. 
This thermal soft (TS) component is presumably emitted from the
whole visible NS surface.
To fit the spectrum at intermediate energies,
one more component is required
(see 
Fig.\ \ref{0656spec1}).
Very likely, this is a thermal hard (TH) component
with temperature $T^\infty_{\rm h}\simeq 1.65$ MK
(as given by the blackbody model)
emitted from pulsar's polar caps of radius 
$R_{\rm pc}=R^\infty_{\rm h}\simeq 1.0\, (d/500\, {\rm pc})$ km heated by
relativistic particles streaming down from the pulsar acceleration zones.
The bolometric luminosity of the TS component is
$L_{\rm bol,s}\simeq 5.5\times 10^{32}\, (d/500\, {\rm pc})^2$ erg s$^{-1}$,
vs.\ $L_{\rm bol,h}\simeq 4.9\times 10^{31}\, 
(d/500\, {\rm pc})^2$ erg s$^{-1}$
for the TH
component.
The luminosity of the nonthermal component is
$L_{\rm x}\simeq 5.3\times 10^{30}\, (d/500\, {\rm pc})^2$ erg s$^{-1}$
in the 0.1--10.0 keV range. This three-component model
fits the data from IR through X-ray energies
(Fig.\ \ref{0656spec2}).

Using the magnetic
hydrogen atmosphere models (ZP02),
instead of blackbodies, for the thermal
components gives formally acceptable fits, with effective temperatures 
lower by a factor of about 2 (e.g., $T_{\rm eff}^\infty \simeq 0.4$--0.5 MK for
the TS component). However, these fits imply either very small distances,
$\sim 70$--130 pc for $R=10$ km, or very large radii, $R_{\rm s}\sim 40$--70 km
for the TS component, at $d=500$ pc. Therefore, applicability of these 
models to PSR 0656+14 is questionable.

The multicomponent structure of the pulsar X-ray
radiation is also seen from
the 
behavior 
of the pulse profiles 
shown
in Figure \ref{0656lc}.
The shape of the light curves is energy
dependent, with a significant phase shift ($\sim 0.2$ of the period)
at 0.5--0.7~keV.
The source pulsed fraction $f_{\rm p}$
shows a complex behavior: it decreases with
increasing energy from about 20\% at $E\sim 0.1$--0.3 keV down to
about 7\% at $E$ around 1 keV and increases again to $f_{\rm p}\simeq 35\%$
at $E\gapr 1$ keV. The X-ray peak lags the radio peak by 0.2--0.4 of the period,
depending on energy
(Fig.\ \ref{0656lc}).

To establish the origin of the peak
and the parameters of the X-ray radiation,
one should
analyze the spectra and the light curves together,
with the aid of realistic models for X-ray emission.
However,
the spectra of the pulsar observed with the
pre-\cha\ X-ray observatories, and even the \cha\ ACIS-S spectrum
observed in CC
mode, potentially have a
contaminating component --- synchrotron
radiation from a PWN of a small angular size
around the pulsar,
which could not be resolved in those observations.

A hint on extended emission around PSR 0656+14
is seen in the zero-order LETG/HRC-S image 
(Marshall \& Schulz 2002),
but that image is strongly contaminated by the diffraction spikes.
The pulsar was also observed in a 5 ks exposure with ACIS-S3
in Timed Exposure mode (December 2001) providing
the data for spatial analysis of the X-ray emission in 
the vicinity of the pulsar.
The analysis 
reveals a $3''$--$4''$ extended structure
around the piled-up image of the pulsar --- first strong indication
on PWN around a middle-aged pulsar (see Fig.\ \ref{0656img}).
The pulsar's image, confined within a small,
slightly elongated area of about $1''$ diameter 
(at the center of Fig.\ \ref{0656img}) is surrounded by
a putative PWN of rather complicated morphology.
The estimated luminosity of the PWN
is $L_{\rm x,pwn}\sim 2\times 10^{31}\, (d/500\, {\rm pc})^2$~erg~s$^{-1}$
in the 0.1--10.0 keV range. 
A considerable fraction of the PWN radiation
can contaminate the pulsar's radiation observed
with the pre-\cha\ observatories and even with ACIS in
CC
mode. Therefore, to unambiguously separate
the PWN contribution,
one should observe the pulsar and its surroundings in high-resolution imaging
mode for longer time.  
The above estimates of the pulsar spectral parameters should be
considered with caution until the properties of the putative PWN
are studied and its radiation is separated from that of the pulsar.

\subsubsection{PSR B1055--52}
Analysis of the \cha\ data on PSR B1055--52, taken in January 2000
with ACIS-S  
in CC 
mode
(42 ks exposure), yields results very similar to those
obtained for PSR B0656+14 (see Sanwal et al. 2002c for details).
First, the ACIS spectrum, combined with that obtained with the \ros\ PSPC,
requires a three-component model (Fig.\ \ref{1055xraysp}).
The high-energy tail of the spectrum ($E\gapr 2$ keV) is
best described with a PL with $\gamma = 1.66$;
such a component smoothly connects optical ({\sl HST} FOC) and
gamma-ray ({\sl CGRO} EGRET) data (Fig.\ \ref{1055spec}).
The spectrum at lower energies fits well with two thermal (blackbody)
components, soft and hard.
The best-fit model parameters to the phase-integrated spectrum are
$T^\infty_{\rm s}=0.84$ MK 
and $R^\infty_{\rm s}=12.0\, (d/1\, {\rm kpc})$ km for the soft
thermal component, 
$T^\infty_{\rm h}=1.62 $ MK
and $R^\infty_{\rm h}=0.8\, (d/1\, {\rm kpc})$ km 
for the hard thermal component,
$n_{\rm H}=1.3\times 10^{20}$ cm$^{-2}$. 

The energy-resolved light curves of the pulsar derived from 
the ACIS data reveal energy-dependent variations in
the pulsed fraction, the phase of the pulse, and 
the pulse shape (Fig.\ \ref{1055lc}). 
In particular, the pulsed fraction
increases from $\sim 15\%$ at about 0.3--0.4 keV
to $\sim 50\%$ at $\gapr 1.0$ keV.
For this pulsar, the X-ray pulse at $E\gapr 0.3$ keV is in phase
with the main radio pulse, contrary to PSR B0656+14.

The phase-resolved spectroscopy shows small variations in
the parameters of the soft thermal component, whereas 
the emitting area for the hard thermal component varies
significantly with the rotational phase ($R^\infty_{\rm h}$
changes between about 0.5 and 1.5 km), which can 
be interpreted as changing 
projection of the
hot spot area.

Thus, we see remarkable similarity in the X-ray properties of the two
middle-aged pulsars, B0656+14 and B1055--52. Interestingly, the temperature
of the soft thermal component is approximately the same for these pulsars,
although the characteristic age, $\tau_c=P/2\dot{P}$, is a factor of 5
larger for B1055--52. A possible explanation of this fact is that these
NSs have different properties (e.g., the mass of B1055--52 is lower ---
see D.G.\ Yakovlev et al., this volume). An alternative explanation is that
the true age of a pulsar can be quite different
from its characteristic age (see below), so that one should be very
cautious comparing the NS cooling models with the observations of pulsars.

\begin{figure}
\centerline{\psfig{file=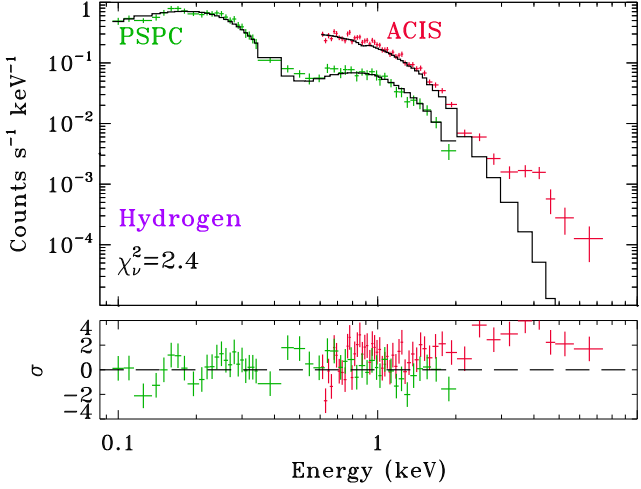,height=7cm,clip=} }
\caption{
Two-temperature hydrogen polar cap
model fit to the combined ACIS and PSPC data
on PSR J0437--4715.
}
\label{0437spec1}
\end{figure}

\begin{figure}
\centerline{\psfig{file=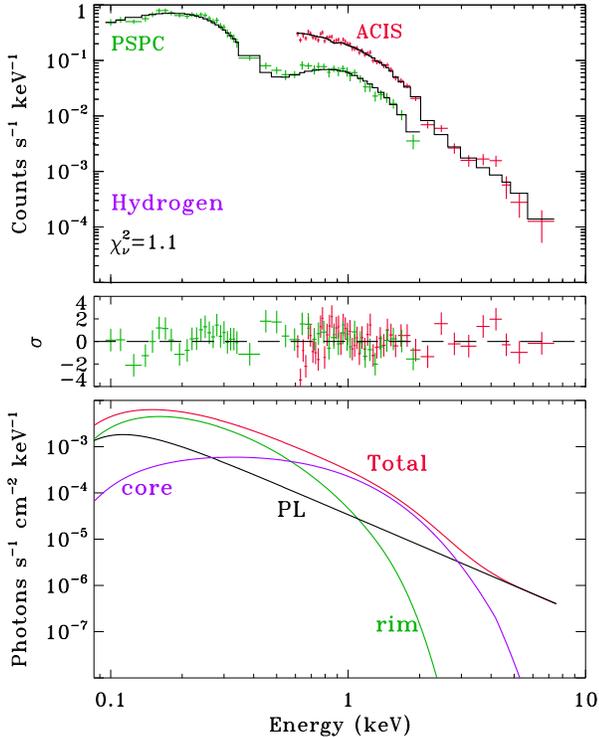,height=10cm,clip=} }
\caption{
Power-law plus two-temperature hydrogen polar cap
model fit to  the combined ACIS and PSPC spectrum of PSR J0437--4715.
}
\label{0437spec2}
\end{figure}

\begin{figure}
\centerline{\psfig{file=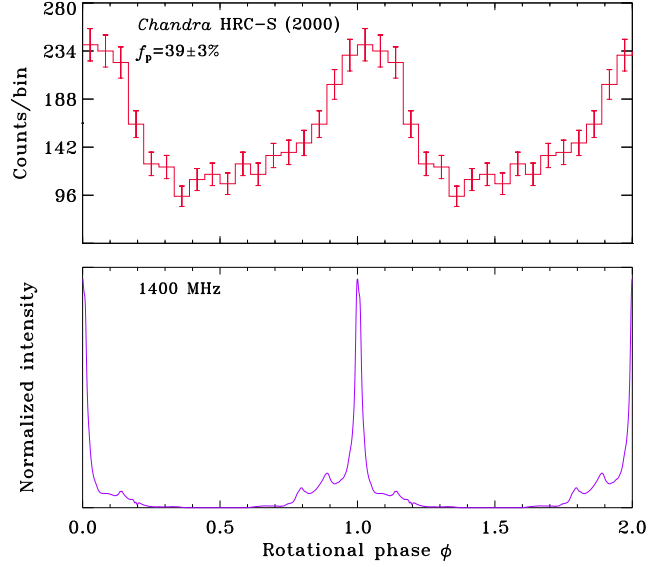,height=8cm,clip=} }
\caption{
Pulse profiles of PSR 0437--4715 from the \cha\ and radio
observations.
}
\label{0437lc}
\end{figure}

\subsection{Millisecond pulsar J0437--4715}
Millisecond (recycled) radio pulsars are distinguished from ordinary pulsars
by their very short and stable periods,
$P\la 10$~ms, $\dot{P}\sim 10^{-20}$~s~s$^{-1}$.
They are presumably very old objects, with spin-down ages
$\tau\sim 10^9$--$10^{10}$~yr and low magnetic fields,
$B\sim 10^8$--$10^{10}$~G.
Similar to ordinary pulsars, a millisecond pulsar can emit
nonthermal X-rays from its magnetosphere.
Additionally, thermal X-rays can be emitted from pulsars'
polar caps
heated up to X-ray
temperatures by relativistic particles
generated in the pulsar acceleration zones.

At a distance $d\simeq140$~pc,
PSR J0437--4715 ($P=5.75$ ms) is the nearest and
brightest millisecond pulsar known
at both radio and X-ray wavelengths.
It is in a 5.74~d binary
orbit with a low-mass white dwarf companion.
Optical observations in H$_\alpha$ have revealed a bow-shock PWN,
caused by the supersonic motion of the pulsar through
the interstellar medium,
with the bow-shock apex at about $10''$ south-east of the pulsar,
in the direction of the pulsar's proper motion.
Its X-ray radiation was discovered with \ros\ (Becker \& Tr\"umper 1993).
The pulsar was also observed with {\sl EUVE} 
(Halpern, Martin, \& Marshall 1996).
The 
pre-\cha\ data, 
obtained in a narrow energy range of 0.1--2.0 keV, can
be well fitted with a PL with photon index $\gamma=2.2$--2.5,
or with a polar cap model with a nonuniform temperature 
distribution (Zavlin \& Pavlov 1998 [ZP98]). To discriminate between these
two competing models, the pulsar was observed with \cha\ ACIS
(May 2000; 26 ks exposure; see Zavlin et al.\ 2002).

The source spectrum
prevails over background at energies up to 7~keV and reveals no
significant spectral features.
No structures which could be associated with
the bow shock observed in H$_\alpha$ are found (see Fig.~\ref{0437img}).

The combined analysis of the ACIS and \ros\ PSPC data 
rules out a single PL model
(${\rm min}~\chi^2_\nu=3.8$), indicating that more
components are needed to get a satisfactory combined fit.

A two-component model, PL plus two polar caps of uniform temperature,
provides an acceptable fit to the combined data.
However, comparison  with the {\sl EUVE} data shows that this
model is only marginally
acceptable. 
In addition, the inferred values of hydrogen column density
exceed the values 
$n_{\rm H}=(1$--$3)\times 10^{19}$ cm$^{-2}$ derived
from independent measurements of the interstellar absorption 
towards objects in the vicinity of the pulsar (see ZP98
for more details).

ZP98 suggested that the temperature distribution in a
polar cap should be nonuniform
because the heat, released by decelerating magnetospheric particles
in subphotospheric layers, propagates along the surface out from the small
polar areas, where the energy is deposited, and heats
a larger area. They used a two-zone approximation for the polar cap
temperature distribution, a hot ``core'' plus a colder ``rim''.
When applied to the ACIS data, this model shows that the observed spectrum
significantly exceeds the model at $E>2$~keV, i.e., 
one more component is required (Fig.\ \ref{0437spec1}).
A natural choice
for the additional component is a PL.  Adding the nonthermal component greatly
improves the fit (Fig.\ \ref{0437spec2}).
At fixed $n_{\rm H}=2\times 10^{19}$ cm$^{-2}$, the best fitting
parameters are
$T_{\rm core}=2.1$ MK,
$T_{\rm rim}=0.54$ MK,
$R_{\rm core}=0.12$ km,
$R_{\rm rim}=2.0$ km,
and $\gamma=2.2$.
The bolometric luminosity of two
polar caps is
$L_{\rm bol}=2.3
\times 10^{30}$~erg~s$^{-1}\sim 
0.6\times 10^{-3}\dot{E}$,
and the PL luminosity in the 0.1--10~keV band is
$L_{\rm x}=0.7
\times 10^{30}$~erg~s$^{-1}$
$\sim 0.2\times 10^{-3}\dot{E}$.
This model is in good agreement with the {\sl EUVE} data.

\begin{table*}
\begin{center}
\begin{tabular}{c|cc|c|ccc|ccc}
\hline
 Object & Host SNR & Age & Period &
 $T^\infty_{\rm bb}$ & $R^\infty_{\rm bb}$ & $L^{\infty}_{\rm bol,bb}$ &
 $T_{\rm eff}^\infty$ & $R^\infty$ & $L_{\rm bol}^\infty$ \\
 &  &  kyr & & MK & km & $10^{33}$erg s$^{-1}$ &
      MK & km & $10^{33}$erg s$^{-1}$   \\
\hline
J2323+5848  & Cas A       & 0.32  & ...     & 5.7 & 0.5     & 1.6    &
3.5 & 1   & 2      \\
J0852--4617 & G266.1--1.2 & $\sim$1 & 301~ms [?] & 4.6 & 0.3     & 0.3    &
3.1 & 1.5 & 0.8    \\
J1617--5102 & RCW~103     & 1--3   & 6.6~hr [?] & 4.6--7.0 & 0.2--1.6 & 0.5--30
 & 3.5 & 1--8 & 1--60  \\
J0821--4300 & Pup A       & 1--3  & ...   & 4.4 & 1.4 & 4.2 &
2.0 &  10  & 8     \\
J1210--5226 & G296.5+10.0 & 3--20 & 424~ms   & 2.9 & 1.6  & 1.3 &
1.6 & 11 & 1 \\
\hline\hline
\end{tabular}
\caption{Compact central objects in SNRs with best-fit parameters
for blackbody (columns 5--7) and hydrogen NS atmosphere
models (columns 8--10).
The periods marked with ``[?]'' require confirmation.
More details on these objects
can be found in Pavlov et al.\ (2002a).
}
\end{center}
\end{table*}

In the polar cap model with
temperature decreasing outwards
from the cap center,
the thermal component 
dominates between 0.1
keV and 2.5 keV, providing about 75\% of the X-ray flux, while the PL component
dominates outside this band.  
An important point here is that this model requires
the thermal radiation to be emitted
from a hydrogen (or helium) NS atmosphere. The high spectral resolution of the
ACIS data rules out an atmosphere comprised of heavier chemical elements.

The pulsar was also observed with the HRC-S detector (October 2000;
20 ks exposure) in the timing
mode that provides accurate timing results.
Figure \ref{0437lc} presents the 
HRC-S 
and radio pulse profiles.
The phases of the X-ray and radio pulses are found to be
virtually the same --- their difference
(0.3\% of the pulsar's period) is smaller than the absolute timing
accuracy of \cha.
This implies that, if the main contribution to 
the HRC-S band is due to the thermal polar cap
radiation, then the radio emission
must be generated close to
the NS surface --- e.g., the time difference of  $<0.1$ ms between the
X-ray and radio phases corresponds to a distance of $<30$ km, much smaller
than the light cylinder radius of 275 km.

\begin{figure}
\centerline{\psfig{file=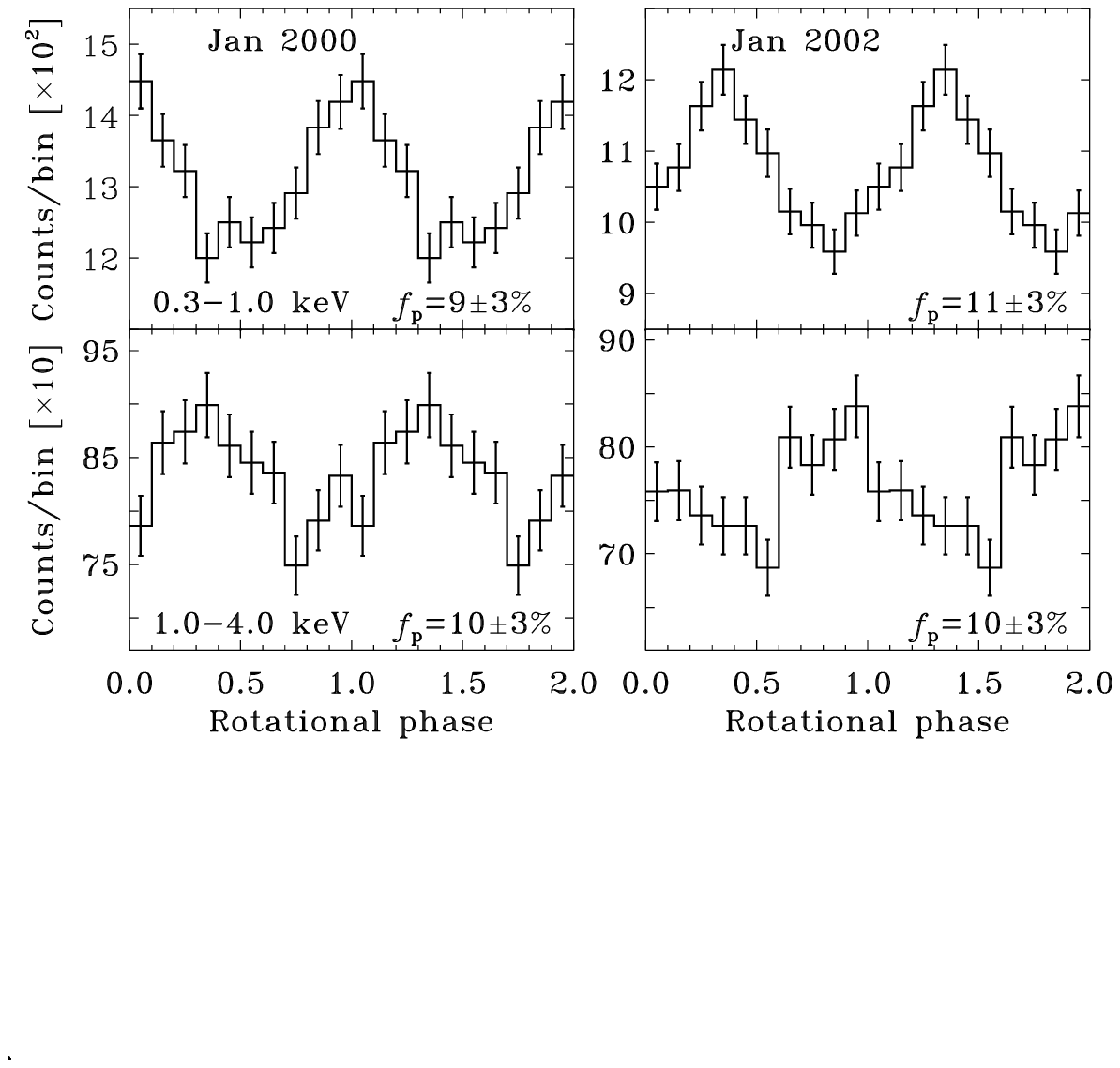,height=8cm,clip=} }
\caption{
Light curves of CXO~J1210--5226 from two \cha\ observations.
}
\label{1207lc}
\end{figure}

\begin{figure*}[ht]
\centerline{\psfig{file=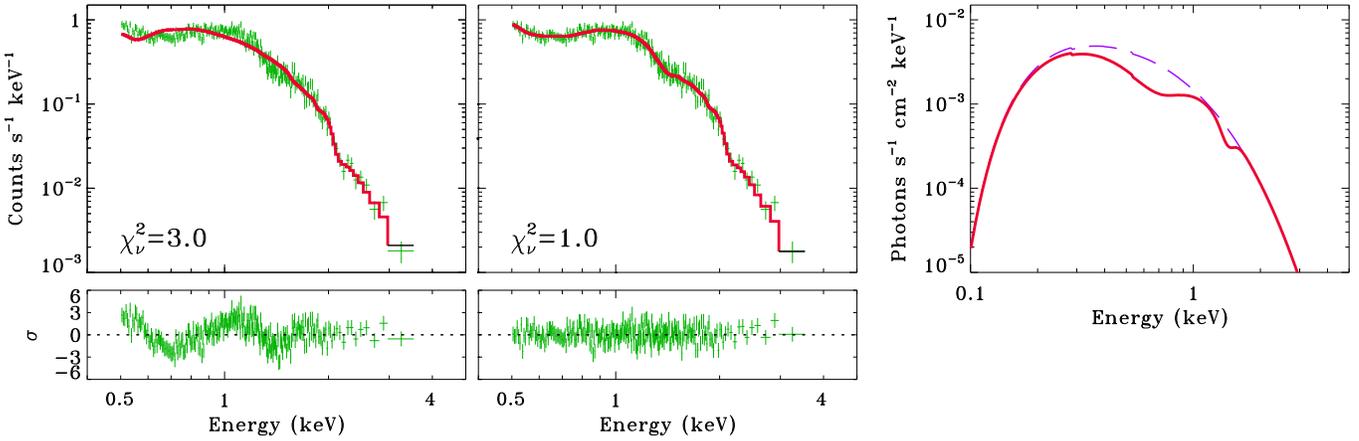,height=8cm,clip=} }
\caption{
{\sl Left:} Fit to the \cha\ spectrum of J1210--5226 
with a blackbody
model. {\sl Middle:} The same model with two absorption
lines added (at energies of 0.7~keV and 1.4~keV).
Notice the significant improvement in the fit quality.
{\sl Right:} The two-line model for the photon flux (see text).
}
\label{1207spec}
\end{figure*}

\section{Compact central objects in supernova remnants}
\cha\ has observed five CCOs.
Their most important characteristics are listed in Table~1.
Although some properties of CCOs
(in particular, their spectra) are very similar to each other,
it does not necessary mean that they represent a uniform class of objects.
The most outstanding 
is J1617--5102 in RCW 103,
with its highly variable flux 
(time scales hours to months), putative 6.6 hr period,
and a possible optical counterpart ($J\simeq 22.3$). 
One can presume that this source is not a
truly isolated NS (i.e., it is powered by accretion, at least in
its high state), although it is not a standard accretion-powered
pulsar.
The other four CCOs have shown neither long-term variabilities nor
indications of binarity. If, based on the observed spectra,
 we adopt a plausible hypothesis that
their radiation is thermal and assume that it can be adequately described
by the same spectral model (e.g., BB or hydrogen 
NS atmosphere),
then we have to conclude that the emitting area is growing with age.
If 
the radiation emerges from a hydrogen
atmosphere, then the effective radius
is consistent with a NS radius for two oldest
CCOs, J0821--4300 and J1210--5226,
being smaller for two youngest ones, J2323+5848 and J0852--4617.
One might assume that 
the latter two are black holes, not NSs,
but then one would have to invoke accretion as the energy source,
which does not look consistent with the lack of variability.

A recent review on CCOs, including the \cha\ results, has been presented
by Pavlov et al.\ (2002a). Therefore, here we will describe only the most
recent results on the best-investigated CCO, J1210--5226
(a.k.a. 1E 1207.4--5209), the first isolated NS
which shows spectral lines.

\subsection{J1210--5226: the puzzling pulsar in G296.5--10}
This 
radio-quiet X-ray
source at the center of G296.5+10.0 (a.k.a. PKS 1209--51/52)
 was discovered
by Helfand \& Becker (1984).
Mereghetti, Bignami, \& Caraveo (1996)
and Vasisht et al.\ (1997)  fitted the \ros\ and \asc\ spectra of
J1210--5226
with a blackbody  model of $T^\infty_{\rm bb}\simeq 3.0$~MK 
from an area with radius
$R^\infty_{\rm bb}\simeq 1.6\, (d/2\, {\rm kpc})$~km.
Zavlin, Pavlov \& Tr\"umper (1998) interpreted
the observed spectra as emitted from a light-element (hydrogen or helium)
atmosphere.  For a NS of mass $1.4\, M_\odot$ and radius 10~km, they
obtained a NS surface temperature $T^\infty_{\rm eff}=(1.4$--1.9)~MK and a
distance of 1.6--3.3 kpc,
consistent with the distance to the SNR,
$2.1^{+1.8}_{-0.8}$~kpc (Giacani et al.\ 2000).

Zavlin et al.\ (2000) observed J1210--5226 with \cha\ 
ACIS-S 
in CC 
mode in January 2000 (32 ks exposure)
and discovered a period of about 424 ms, which proved that the
source is a NS.
Second \cha\ observation with the same observational setup
(January 2002; 30 ks)
provided an estimate of the
period derivative,  $\dot{P} \sim (0.7$--$3) \times 10^{-14}$ s s$^{-1}$
(Pavlov et al.\ 2002b).  This estimate implies that the characteristic age
of the NS, $\tau_c \sim 200$--1600 kyr, is much larger than the 3--20 kyr age
of the SNR (Roger et al.\ 1998), while the ``canonical'' magnetic field,
$B\equiv 3.2\times 10^{19} (P \dot{P})^{1/2}\, {\rm G} = (2$--$4)
\times 10^{12}$  G, is typical for a radio pulsar. 
These observations revealed smooth pulsations of the X-ray flux
from J1210--5226, with pulse shape depending on photon energy
(Fig.\ \ref{1207lc}).

The analysis of the ACIS spectra (Sanwal et al.\ 2002b)
shows that continuum models 
(e.g., blackbody, power-law, fully ionized NS atmosphere)
fail to fit the source 
spectrum, which
deviates very significantly from any of the models,
revealing two absorption features near
0.7 keV and 1.4 keV 
(Fig.\ \ref{1207spec}).
Including absorption lines in the spectral model improves the quality of the
fit substantially.
Assuming Gaussian profiles for
the absorption coefficients, we obtain the central energies $0.73\pm 0.01$ 
and $1.40\pm 0.01$ keV, equivalent widths $108\pm 15$ and $112\pm 24$ eV,
and central optical depths $0.40\pm0.06$ and $0.43\pm0.06$, respectively.

Attempts to explain the absorption features
as caused by the intervening interstellar
(or circumstellar) material lead to huge overabundance for some elements.
Therefore, the observed lines are most likely  intrinsic to the NS.

One could assume that the observed lines are cyclotron lines produced
in a strongly ionized NS atmosphere. If one assumes these are electron
cyclotron lines,  the features might be interpreted as the fundamental
and the first harmonic of the electron cyclotron energy $E_{ce}=1.16 B_{11}$
keV in a
magnetic
field $B_{11}\equiv B/(10^{11}\, {\rm G})\sim 0.6\, (1+z)$
($z$ is the neutron star redshift parameter),
or as two fundamentals emitted from two regions with different magnetic
fields, $B_{11}=0.6\,(1+z)$
and $1.2\,(1+z)$,
broadened by the
radiative transfer effects and/or nonuniformity of the magnetic field.
However, it is difficult to reconcile this interpretation with the
expected strength of the surface magnetic field $B> 3\times 10^{12}$~G.
In addition, the oscillator strength
of the first harmonic is  smaller than that of the fundamental by a factor
of $\sim E_{ce}/(m_ec^2) \sim 2\times 10^{-3}$ (at $E_{ce}\gg kT$),
so that it is hard to explain why the 1.4 keV feature is as strong as
the 0.7 keV feature if we assume the two lines are associated with the
same magnetic field.

\begin{table*}
\begin{center}
\begin{tabular}{c|ccccc|c}
\hline
 Object & $T^\infty_{\rm bb}$ & $f_{\rm x}$$^a$
                &  $n_{\rm H}$ & Period & $B$ & Ref. \\
 & MK &
                        & $10^{20}$ cm$^{-2}$ & s &  mag & \\
\hline
RX~J1856.5--3754 & 0.7 & 14 & 1.3 & ...  & 25.8      &  1 \\
RX~J0720.4--3125 & 0.9 & 11 & 1.1 & 8.37 & 26.6      &  2,3 \\
RX~J1605.3+3249  & 1.1 & 9 & 1.1 & ...  & $>25$     &   4 \\
RX~J0806.4--4123 & 0.9 & 3  & 2.5 & 11.37[?] & $>24$ & 5,6 \\
RX~J0420.0--5022 & 0.7 & 1  & 1.7 & 22.69[?] & $>25.2$ &  7 \\
RX~J1308.8+2127  & 1.4 & 5 & 2.1 & 5.16  & $>26$ &  8,9   \\
RX~J2143.0+0654  & 1.1 & 9 & 4.6 & ...   & $>22.8^{\rm b}$ & 10 \\
\hline\hline
\end{tabular}
\caption{
Main properties of the seven ``dim'' NSs discovered with \ros.
The spectral parameters are obtained from
blackbody fits
to the \ros\ PSPC data.
The periods marked with ``[?]'' require confirmation.
\newline
Refs.: 1 - Walter et al.\ (1996); 2 - Haberl et al.~(1997);
3 - Kulkarni \& van Kerkwijk (1998); 4 - Motch et al.~(1999);
5 - Haberl et al.~(1998); 6 - Haberl \& Zavlin (2002);
7 - Haberl et al.~(1999); 8 - Schwope et al.~(1999);
9 - Hambaryan et al.~(2001); 10 - Zampieri et al.~(2001).
\newline
\footnotesize{$^{\rm a}$~Absorbed flux in 0.1--2.0 keV, in units of
$10^{-12}$ erg cm$^{-2}$ s$^{-1}$.}
\newline
\footnotesize{$^{\rm b}$~$R$ magnitude.}
}
\end{center}
\end{table*}

One can assume that the spectral features are associated
with ion cyclotron energies, $E_{ci}=0.63 (Z/A) B_{14}$ keV,
where $Z$ and $A$ are the ion's charge and mass numbers.
The surface
magnetic field needed for this interpretation
is $\ga 10^{14}$ G, much
larger than the conventional magnetic field inferred for a centered
dipole from the $P$, $\dot{P}$ measurements.
Such a strong {\em surface} magnetic field can be consistent with the
magnetic moment $\mu = 3\times 10^{30}$ G cm$^3$  if the magnetic
dipole  is  off-centered to about 3 km below the surface.
However, the ion-cyclotron interpretation of the observed features
is not straightforward even in this case
(see Sanwal et al.~2002b for more details).

Thus, the most viable
possibility is that the observed lines are  atomic lines
formed in the NS atmosphere.
The 
features
cannot be explained as emerging from a hydrogen atmosphere
because, at any magnetic field and any reasonable gravitational
redshift, there is
not a pair of strong hydrogen
spectral lines whose energies would match the observed ones.
Pavlov et al.\ (2002c) suggest that these are the strongest lines
of once-ionized helium in a superstrong magnetic field, (1.4--1.7)$\times
10^{14}$ G. In this interpretation, the observed line energies
correspond to the gravitational redshift  $z=0.12$--0.23, which corresponds to
$R/M=8.8$--14.2 km~$M_\odot^{-1}$. 
To confirm this result,
deep observations with high spectral resolution are needed.

\section{``Dim'' isolated neutron stars}

\begin{figure}[!ht]
\centerline{\psfig{file=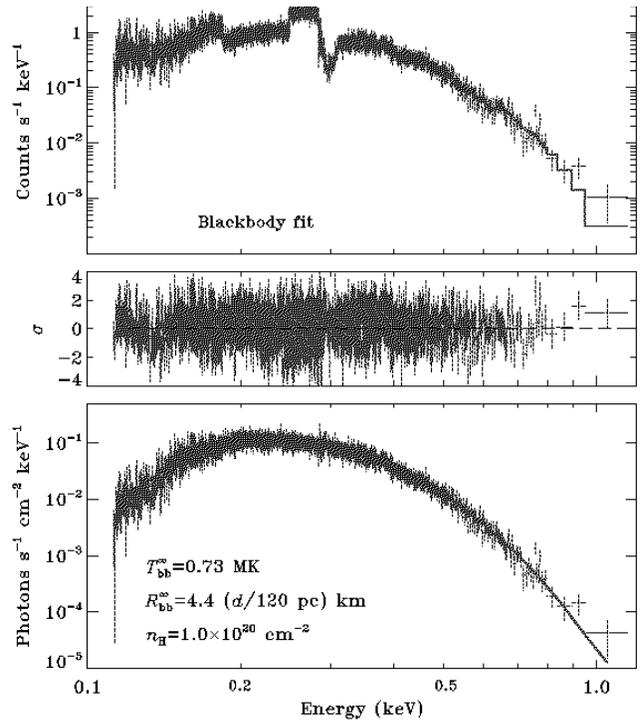,height=10cm,clip=} }
\caption{
\cha\ LETG/HRC-S spectrum of RX~J1856.5--3754 collected in a 500 ks
exposure. The red curves represent
the best blackbody fit to
the observational data (the model parameters are given in the
bottom panel).
}
\label{1856spec1}
\end{figure}

\begin{figure}[!hb]
\centerline{\psfig{file=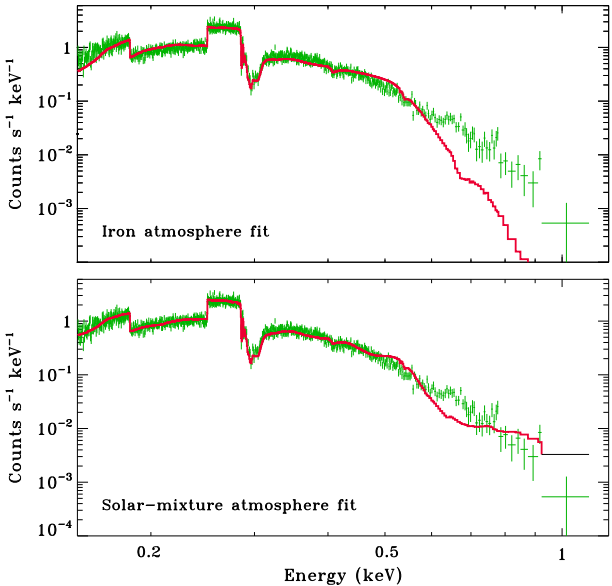,height=8cm,clip=} }
\caption{
Fits to the LETG spectrum of RX~J1856.5--3754 
to NS atmosphere models of heavy-element compositions
assuming that the NS is a fast rotator (the models
are from ZP02).
}
\label{1856spec2}
\end{figure}

\begin{figure}[!ht]
\centerline{\psfig{file=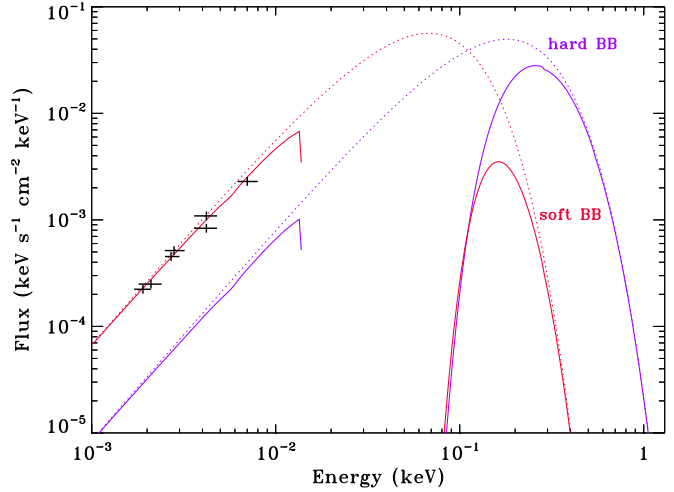,height=8cm,clip=} }
\caption{Spectral flux of RX~J1856.5--3754 as given
by the best blackbody fit to the LETG data (blue curves, marked ``hard BB''),
with $T^\infty_{\rm bb}=0.73$ MK and
$R^\infty_{\rm bb}=4.4\, (d/120\, {\rm pc})$ km.
The red curves give an example of a blackbody fit with
$T^\infty_{\rm bb,s}=0.30$ MK and
$R^\infty_{\rm bb,s}=18.2\, (d/120\, {\rm pc})$ km (``soft BB'')
to the optical data (crosses).
Solid curves show the absorbed spectra 
at $n_{\rm H}=1.0\times 10^{20}$ cm$^{-2}$
(dots are the unabsorbed models).
}
\label{1856spec3}
\end{figure}

\ros\ discovered seven objects
which are often called ``dim'' or ``truly isolated''
NSs. These sources, listed in Table~2,
are characterized
by 
blackbody-like spectra
with temperatures around 1 MK.
Low values
of hydrogen column densities, $n_{\rm H}\sim 10^{20}$ cm$^{-2}$,
suggest small distances to these objects.
Their X-ray fluxes, 
$f_{\rm x}\sim 10^{-12}$--$10^{-11}$ erg s$^{-1}$ cm$^{-2}$,
correspond to rather low
(bolometric) luminosities, $L\sim 10^{30}-10^{31}$ erg s$^{-1}$
for a fiducial distance of 100 pc. The fluxes
 do not show variations on either short
(minutes-hours) or long (months-years) time scales, and they are much higher
than the optical fluxes,
$f_{\rm x}/f_{\rm opt}\ga 10^3$.
Such properties strongly suggest that these objects are isolated, nearby
NSs. Rotational periods in J0720.4--3125
and J1308.8+2127 were discovered with \ros\ and \cha, respectively
(see refs.\ in Table~2). 
There are also indications on periodicities in
J0420.0--5022 and J0806.4--4123, although those periods
are to be confirmed yet.
The source powering 
the thermal radiation of these NSs 
has been a 
matter of debate (e.g., Treves et al.\ 2000). 
Two main options under consideration are the internal heat of a
cooling NS
and accretion from 
the interstellar medium (ISM). 
A standard estimate on the luminosity 
powered by the Bondi-Hoyle accretion, 
$L_{\rm x}\sim 2\times 10^{31} n v^{-3}_{10}$ erg s$^{-1}$ (here
$n$ is the number density of the ISM in cm$^{-3}$, 
$v_{10}$ is the relative NS velocity in units of 10 km s$^{-1}$),
requires rather low NS velocities.
Recent estimates on the distance to J1856.4--3754 and its proper
motion (see 
below) imply a transverse velocity
$v_\perp \simeq 185$ km s$^{-1}$. 
Therefore, accretion cannot be
the energy source for this object
unless
the star moves in 
a surprisingly dense medium (a cloud?) with $n\sim 10^2$ cm$^{-3}$.
First estimates of period derivatives 
for J0720.4--3125 (Zane et al.~2002; Kaplan et al.~2002b)
and J1308.8+2127 (Hambaryan et al.~2002) imply surface
magnetic fields\footnote{
We note that these results are preliminary and require 
further confirmation.} $\sim 10^{13}-10^{14}$ G,
which favors models invoking young (or middle-aged) 
cooling NSs rather than accreting NSs.

Below we present results from \cha\ observations of the most famous 
(and best-investigated) objects
of this class, J1856.5--3754 and J0720.4--3125.

\subsection{RX~J1856.5--3754}
This object (we will call it J1856 throughout this Section)
was first observed with \cha\ in March 2000 in a 55 ks exposure with
the HRC-S/LETG instrument. A detailed analysis of this observation
is given by Burwitz et al.\ (2001), who reported
no detection of any significant features in the high-resolution spectrum.
They also concluded that the combined X-ray and optical data
rule out all the available NS atmosphere models: light-element models
overpredict the actual optical fluxes
(as first estimated by Pavlov et al.\ 1996),
whereas heavy-element models
do not fit the LETG spectrum.
In October 2001
this source was observed for 450 ks,
with the same instrument.
However, even this very deep observation
revealed no spectral features.
The spectrum excellently fits a blackbody model of 
$T^\infty_{\rm bb}=0.73$ MK and
$R^\infty_{\rm bb}=4.4\, (d/120\, {\rm pc})$ km,\footnote{
The distance to J1856,
$d\simeq 120$ pc,  has been recently established
by Kaplan et al.~(2002a) and Walter \& Lattimer (2002).}
with the hydrogen
column density of $n_{\rm H}=1.0\times 10^{20}$ cm$^{-2}$
(Fig.\ \ref{1856spec1}; see also Drake et al.~2002).

All attempts to find pulsations from J1856 have been unsuccessful
so far (e.g., Ransom et al.\ 2002).
However, since time resolution of both the \ros\ and \cha\ data on J1856
was not sufficient to search  for periods shorter than $\approx 10$ ms,
one could assume that the period of J1856 is shorter than
that.
Such a hypothesis might also explain the featureless spectrum of J1856 ---
spectral lines in radiation of a fast-rotating NS
can be 
smeared out
because of the Doppler effect (rotational broadening; see ZP02 
for more details). 
However, 
atmosphere models for a fast-rotating NS fail to
fit the detected spectrum
because of strong deviations 
at energies around most prominent absorption edges 
(see Fig.\ \ref{1856spec2} and Fig.~12 in ZP02).

\begin{figure}[!ht]
\centerline{\psfig{file=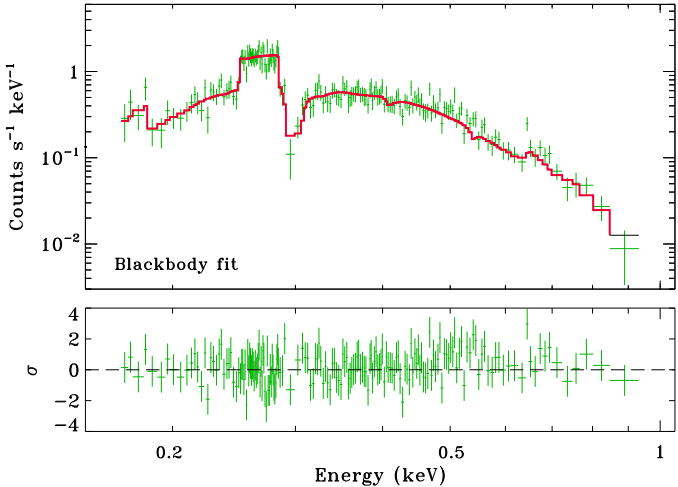,height=7cm,clip=} }
\caption{
\cha\ LETG spectrum of RX~J0720.4--3125.
The red curve shows the best blackbody fit to
the observational data (model parameters are given in the text).
}
\label{0720spec1}
\end{figure}

Thus, we have to admit that the blackbody spectrum is the best model
for the X-ray spectrum of J1856.
In principle, such a blackbody
spectrum could
mimic the actual spectrum in the relatively short band of 0.15--1.0 keV
if the NS surface is in a condensed form (according to Lai \& Salpeter 1997,
this 
can happen if the magnetic field is very strong at the NS surface, 
$B\gapr 10^{14}$ G). However, the radius of the emitting area,
$R^\infty_{\rm bb}\simeq 4$--5 km, 
is too small to be the NS radius.
Moreover, taking into consideration
the optical fluxes detected from J1856
(van Kerkwijk \& Kulkarni 2001a; Pons et al.~2002),
one encounters one more problem --- the extrapolation of the blackbody
model inferred from
the X-ray data into the optical band
falls
well below the actually detected fluxes.
One may assume (Pons et al.\ 2002) that the X-ray
radiation is emitted from a hot area on
the NS surface, 
while the optical radiation
is emitted from the
rest of the surface, with a lower temperature.
The optical spectrum is
well described by the Rayleigh-Jeans law 
(van Kerkwijk \& Kulkarni 2001a) and gives the following
relation 
between the temperature and the radius of the NS surface emitting
the soft blackbody radiation:
$R^\infty_{\rm bb,s}\simeq 
8.3\, (T^\infty_{\rm bb,s}/1~{\rm MK})^{-1/2} (d/100~{\rm pc})$ km
(with allowance for the contribution of the ``hard blackbody'').
On the other hand, 
a 3$\sigma$ upper limit on the temperature of the soft component, 
$T^\infty_{\rm bb,s}<0.39$ MK, found from the X-ray data, 
corresponds to a lower limit for the radius 
$R^\infty_{\rm bb,s} > 16.3$ km for $d=120$ pc.
Figure~\ref{1856spec3} shows an example of the two-blackbody
fit to the optical and X-ray data on J1856.
In this two-temperature picture, it is tempting to interpret the hot area
as a pulsar polar cap. This seems to be 
consistent with the recent discovery
of the H$_\alpha$ 
nebula surrounding J1856 (van Kerkwijk \& Kulkarni 2001b), 
presumably a bow shock in the ambient medium 
created by the supersonic motion of a NS
with a relativistic (pulsar?) wind. 
However, the hot spot is unusually large and luminous
[$L_{\rm bol}^\infty = 4\times 10^{31} (d/120\, {\rm pc})^2$ erg s$^{-1}$]
for a standard pulsar polar cap. This could be explained by a very strong
magnetic field localized in the spot region (i.e., essentially
different from a centered dipole). Such a field can quench the pair cascade
in the wind of primary particles pulled out from the NS surface,
which explains the lack of X-ray/optical synchrotron radiation.  
On the other hand,  the strong field can increase the 
heat flux from the core/crust to the surface. 
The large size of the hot spot,
together with the GR effects (ZP02),
can explain the nondetection of pulsations
(pulsed fraction $f_{\rm p}\la 4\%$
from the \cha\ data), particularly if
the hot spot is close the rotational pole                         
or the direction of the rotational axis is close to the line of sight.
If the new observation of J1856
with {\sl XMM}-Newton
(conducted in April 2002)
reveals pulsations of this objects, it will support 
the two-temperature interpretation and provide 
constraints on the magnetic moment, hot spot position and inclination of the
rotational axis. 

\begin{figure}[!hp]
\centerline{\psfig{file=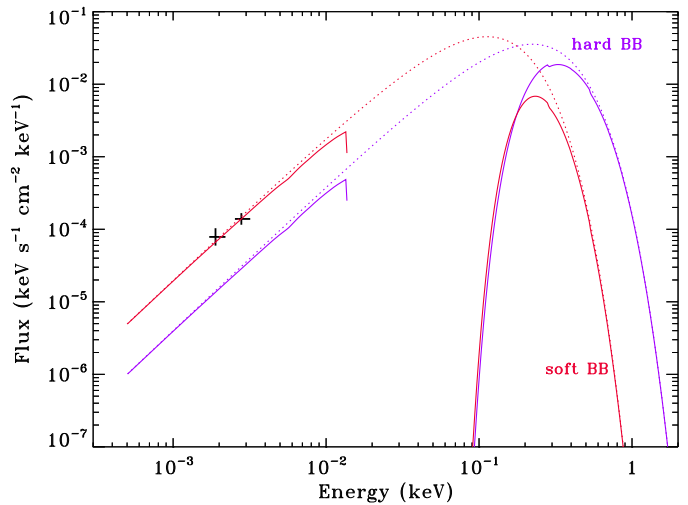,height=8cm,clip=} }
\caption{The same as in Fig.\ \ref{1856spec3},
for RX~J0720--3125.
The parameters for the
model spectra are given in the text. The optical data (crosses) are
from Kulkarni \& van Kerkwijk (1998).
}
\label{0720spec2}
\end{figure}

\subsection{RX~J0720.4--3125}
\cha\ observed this object in February 2000 
for 38 ks
with the HRC-S/LETG instrument.
Fitting various spectral models to these data gives results
remarkably similar to those for J1856.
Although the light-element (hydrogen
and helium) NS atmosphere models provide 
acceptable fits
(but the models are 
somewhat harder than the observed
spectrum at $E>0.75$ keV), the predicted optical fluxes
(e.g., $B=23.7$ and 23.3 for the hydrogen and helium 
nonmagnetic models,
respectively) are much brighter
than 
observed ($B\simeq 26.6$ --- see
Kulkarni \& van Kerkwijk 1998). 
Neither iron nor solar-mixture models fit the data because of
numerous 
features in the model spectra.
The best fit is provided by a blackbody model of
$T^\infty_{\rm bb}=0.92$ MK and a radius of the emitting
area of $R^\infty_{\rm bb}=2.2\, (d/100\, {\rm pc})$ km,
absorbed with the hydrogen column density 
$n_{\rm H}=1.7\times 10^{20}$ cm$^{-2}$ (Fig.\ \ref{0720spec1}). 
However, 
similar to J1856, the best blackbody model
inferred from the X-ray data underpredicts the optical fluxes which
can be interpreted as emitted by a ``soft blackbody'' with
$R^\infty_{\rm bb,s}\simeq 4.3\, 
(T^\infty_{\rm bb,s}/1~{\rm MK})^{-1/2} (d/100\, {\rm pc})$ km.
To obtain an acceptable fit of the two-blackbody model
to the X-ray data, one
should keep $T^\infty_{\rm bb,s}\la 0.50$ MK, that
results in $R^\infty_{\rm bb,s}\ga 6.1\, (d/100\, {\rm pc})$ km.
Figure~\ref{0720spec2} shows an example of the two-blackbody
fit to the optical and X-ray data, with
$T^\infty_{\rm bb,s}=0.45$ MK and 
$R^\infty_{\rm bb,s}=6.4\,  (d/100\, {\rm pc})$ km for the soft component.
If future optical observations provide the distance to J0720.4--3125,
it will be possible to evaluate the NS radius and, based on the X-ray
light curves, estimate the mass-to-radius ratio. 

\section{Concluding remarks}
Because of the limited space, we described the observational results
on only a handful of observed
thermally emitting isolated NSs. We only briefly
mentioned 
AXPs and SGRs, whose X-ray radiation almost certainly contains a thermal
component, with temperatures of 5--10 MK and 
equivalent radii of 1--5 km. The current status of AXPs, including their
comparison with SGRs, is
reviewed by S.\ Mereghetti et al.\ (this volume).
 From the observational perspective, the nature of these enigmatic objects
can hardly be understood without deep multiwavelength observations,
particularly in the IR and hard X-ray bands, which would help elucidate
the origin of the
nonthermal component and separate the thermal component more
precisely.
We did not touch ``elderly'' nearby pulsars (e.g., B1929--10, B0950+08),
mostly because they have not been observed with \cha. 
If future deep observations show that their
X-ray radiation is indeed thermal (emitted from hot, small polar caps),
as suggested by the \ros\ and \asc\ observations, the analysis of
the spectra and light curves will help establish the strength and
geometry of their magnetic fields and the NS mass-to-radius ratio.
We did not mention important non-detections of thermal radiation
from very young 
pulsars, the Crab (Tennant et al.\ 2001; M.C.\ Weisskopf, this volume) 
and the recently discovered PSR J0205+6449
in the SNR 3C\, 58 (Slane et al.\ 2002),
which put upper limits on the NS surface temperature. Finally,
very interesting results on thermal radiation from transiently
accreting NSs in quiescence have been reported recently
(Rutledge et al.\ 2001, and references therein). 
Although these old NSs are not truly
isolated, their radiation in quiescence resembles that of young
isolated NSs because of additional heating due to pycnonuclear
reactions in the compressed accreted material.

The \cha\ observations described above have shown a few important
things. First of all, they left no doubts that thermal emission indeed dominates
in the soft X-ray radiation of quite a few isolated NSs, and the analysis
of this radiation allows one to measure the NS temperatures\footnote{See
Fig.\ 5 in Mereghetti et al.\ (this volume) which demonstrates
current estimates for the temperatures
and luminosities of 22 thermally emitting NSs.} and,
in some cases, the radii or mass-to-radius ratios. 
Particularly important is the conclusion
that the simple picture of a NS with a centered dipole magnetic
field and 
uniform surface temperature
is, most likely,
an oversimplification. We see that whenever the data allow a detailed
analysis, the temperature is not uniform, in both active pulsars
and radio-quiet NSs. Moreover, the radius of X-ray emitting area
is, as a rule, considerably smaller than the ``canonical'' NS radius.
The example of J1210--5226 (Sec.\ 3.1) shows that the characteristic
age ($\tau_c=P/2\dot{P}$) of a pulsar can differ from its true age
by a large factor, and the conventional ``pulsar magnetic field''
can be quite different from the actual magnetic field at the NS surface.
Therefore, any inferences on the properties of NSs and the superdense
matter obtained without taking this into account should be considered
with caution. 
For example, 
adopting the X-ray emitting radius to be the NS radius,
a number of authors hurried to conclude
that NSs are strange quark stars.
Another example is 
the numerous comparisons of the NS cooling curves
with observations of thermal X-rays
from pulsars postulating that the pulsar's characteristic age is the true
age and the empirical temperature of one of the ``thermal
components'' is a single effective 
temperature.
If we want to understand the real objects, 
we should abandon the naive paradigm and answer the
basic questions raised by the observations:
\linebreak
$\bullet$ What is the true explanation of the small, hot areas in many,
if not all, NSs, such as J2323+5848 (CCO of Cas A), RX J1856.5--3754, etc?
If these small heated areas are associated not with (unseen) pulsar
activity, but with superstrong local magnetic
fields, how such fields affect the thermal evolution of NSs?
\linebreak
$\bullet$ What is the relation between the various kinds of NSs --- CCOs,
AXPs, SGRs, active pulsars, dim NSs? For 
instance, why do CCOs, AXPs and SGRs show
similar properties of thermal radiation, being so different in the
other observational manifestations? Can it be that the members of 
different subclasses are intrinsically similar objects at different stages
of their evolution (e.g., AXPs are descendants of CCOs) or viewed
from different directions (e.g., dim NSs are usual middle-aged pulsars
whose pulsar radiation is not seen because of unfavorable geometry)?
\linebreak
$\bullet$
What is the actual temperature distribution over the NS surface?
Does the two-component model for thermal radiation
we have used (thermal soft + thermal hard) describe the
distribution adequately?
\linebreak
$\bullet$ What is the actual mechanism(s) of the surface emission?
Are there any gaseous atmospheres at all? Why does the blackbody model 
describes 
the observed spectra so well in many cases?\\
These are the questions 
to be addressed (and, hopefully, answered) by future
observations of NSs.  

\begin{acknowledgements}
We gratefully acknowledge the support by the Heraeus foundation
and the hospitality of the Physikzentrum Bad Honnef.
The work of GGP and DS was partly supported by NASA grant NAG5-10865
and SAO grants GO2-3089X, GO2-3088X and SP2-2001C.
We thank Marcus Teter and Vadim Burwitz, 
who participated in the analysis of some
of the above-described \cha\ observations. 
\end{acknowledgements}


\begin{thebibliography}{} 

\bibitem{}
	Becker W., Pavlov G.G., 2002, in 
	The Century of Space Science,
	eds.\ J.~Bleeker, J.~Geiss \& M.~Huber (Dordrecht: Kluwer),
	in press
\bibitem{}
        Becker W., Tr\"umper J., 1993, Nature, 365, 528
\bibitem{}
	Becker W., Tr\"umper J., 1997, A\&A, 326, 682  
\bibitem{}
	Brinkmann W., \"Ogelman H., 1987, A\&A, 182, 71
\bibitem{} 
	Burwitz V., Zavlin V.E., Neuh\"auser R., Predehl P.,
	Tr\"umper J., Brinkman A.C., 2001, A\&A, 379, L35
\bibitem{}
	Caraveo P.A., De Luca A., Mignani R.P., Bignami G.F.,
	 2001, ApJ, 561, 930
\bibitem{}
	Fahlman G.G., Gregory P.C., 1981, Nature, 293, 202
\bibitem{} 
	Cheng A.F., Helfand D.J., 1983, ApJ, 271, 271
\bibitem{} 
	C\'ordova F.A., Hjellming R.M., Mason K.O.,
	Middleditch J., 1989, ApJ, 345, 451

\bibitem{} Drake J.J., Marshall H.L., Dreizler S.,
Freeman P.E., Fruscione A., Juda M., Kashyap V., Nicastro F.,
Pease D.O., Wargelin B.J., Werner K., 2002, ApJ, in press

\bibitem{} Giacani E.B., Dubner G.M., Green A.J., Goss W.M., 
Gaensler B.M., 2000, AJ, 119, 281

\bibitem{} Gould D.M., Lyne A.G., 1998, MNRAS, 301, 235

\bibitem{} Haberl F., Zavlin V.E., 2002, A\&A, in press (astro-ph/0205460)

\bibitem{} Haberl F., Motch C., Buckley D.A.H., Zickgraf F.J., Pietsch W., 
1997, A\&A 326, 662

\bibitem{} Haberl F., Motch C., Pietsch W., 1998, Astronom. Nach., 319, 97

\bibitem{} Haberl F., Pietsch W., Motch C., 1999, A\&A, 351, L53

\bibitem{}
	Halpern J.P., Martin C., Marshall H.L., 1996, ApJ, 462, 908

\bibitem{} Hambaryan V., Hasinger G., Schwope A.D., Schulz N.S.,
2002, A\&A, 381, 98

\bibitem{}
	Harding A.K., Strickman M.S., Gwinn C., McCulloch P., Moffet D., 2002,
	ApJ, in press (astro-ph/0205183)

\bibitem{}
	Helfand D.J., Becker R.H., 1984, Nature, 307, 215

\bibitem{}
	Helfand D.J., Gotthelf E.V., Halpern J.P., 2001, ApJ, 556, 380

\bibitem{} Kaplan D.L., van Kerkwijk M.H., Anderson J., 2002a, ApJ, 566, 378

\bibitem{} Kaplan D.L., Kulkarni S.R., van Kerkwijk M.H., Marshall H.K.,
2002b, ApJ, 570, L79

\bibitem{} Kellett B.J., Branduardi-Raymont G., Culhane J.L., Mason I.M.,
 Mason K.O., Whitehouse D.R., 1987, MNRAS, 225, 199

\bibitem{}
	Koptsevich A.B., Pavlov G.G., Zharikov S.V.,
Sokolov V.V., Shibanov Yu.A., Kurt V.G., 2001, A\&A, 370, 1004

\bibitem{} Kouveliotou C., 1995, Ap\&SS, 231, 49

\bibitem{} Kulkarni S.R., van Kerkwijk M.H., 1998, ApJ, 507, L49

\bibitem{} Lai D., Salpeter E.E., 1997, ApJ, 491, 270

\bibitem{}
	Marshall H.L., Shultz N., 2002, ApJ, in press (astro-ph/0203463)

\bibitem{} Mereghetti S., Bignami G.F., Caraveo P.A., 1996, ApJ, 464, 842

\bibitem{} Motch C., Haberl F., Zickgraf F.J., Hasinger G., 
Schwope A.D., 1999, 351, 177

\bibitem{}
	\"Ogelman H., 1995, in The Lives of the Neutron Stars,
	eds.\ M.A.\ Alpar, \"U.\ Kizilo\v{g}lu and J.\ van Paradijs
	NATO ASI Ser.\ (Dordrecht: Kluwer), p.101

\bibitem{} Pavlov G.G., Stringfellow G.S., C\'ordova F.A.,
1996, ApJ, 489, L75

\bibitem{} Pavlov G.G., Zavlin V.E., Tr\"umper J., Neuh\"auser R.,
1996, ApJ, 472, L33

\bibitem{}
	Pavlov G.G., Welty A.D., C\'{o}rdova F.A., 1997, ApJ, 489, L75

\bibitem{}
	Pavlov G.G., Sanwal D., Garmire G.P., Zavlin V.E.,
Burwitz V., Dodson R.G., 2000, AAS Meeting 196, \#37.04 

\bibitem{}
	Pavlov G.G., Kargaltsev O.Y., Sanwal D., Garmire G.P., 2001a, ApJ,
554, L189

\bibitem{}
	Pavlov G.G., Zavlin V.E., Sanwal D., Burwitz V., Garmire G.P.,
   2001b, ApJ, 552, L129

\bibitem{} Pavlov G.G.,  Sanwal D., Garmire G.P., Zavlin V.E., 2002a,
in Neutron Stars in Supernova Remnant, ASP Conf. Series,
eds. P.O. Slane \& B.M. Gaensler, in press (astro-ph/0112322)

\bibitem{} Pavlov G.G., Zavlin V.E., Sanwal D., Tr\"umper J., 2002b,
ApJ, 569, L95

\bibitem{} Pavlov G.G., Sanwal, D., Teter, M.A., Zavlin, V.E, 2002c, 
	AAS Meeting 200, \#80.01

\bibitem{} Pons J.A., Walter, F.M., Lattimer J.M., Prakash M.,
Neuh\"auser R., An P., 2002, ApJ, 564, 981

\bibitem{} Ransom S.M., Gaensler B.M., Slane P.O., 2002, ApJ, 570, L75

\bibitem{}
	Rutledge R.E., Bildsten L., Brown E.F., Pavlov G.G.,
	Zavlin V.E. 2001, ApJ, 559, 1054

\bibitem{}
	Sanwal D., Pavlov G.G., Kargaltsev O.Y., Garmire G.P., Zavlin V.E.,
	Burwitz V., Manchester, R.N., Dodson R., 2002a, in Neutron Stars
	in Supernova Remnants, eds.\ P.O.\ Slane and B.M.\ Gaensler,
	in press (astro-ph/0112164)
\bibitem{} Sanwal D., Pavlov G.G., Zavlin V.E.,
	Teter M.A., 2002b, ApJ, submitted 
\bibitem{}
	Sanwal D., Pavlov G.G., Zavlin V.E.,
	Teter M.A., Tsuruta S., Manchester R.N., 2002c, ApJ, submitted

\bibitem{}
	Slane P.A., Helfand D.J., Murray S.S. 2002, ApJ, 571, L45

\bibitem{}
	Seward F.D., Charles P.A., Smale A.P., 1986, 305, 814 

\bibitem{} Schwope A.D., Hasinger G., Schwarz R.,
Haberl F., Schmidt M., 1999, A\&A, 341, L51 

\bibitem{}
	Tennant A.F., Becker W., Juda M., et al., 2001, ApJ, 554, L173

\bibitem{} Treves A., Turolla R., Zane S., Colpi M., 2000, PASP, 112, 297

\bibitem{}
	Tsuruta S., 1998, Phys. Rep., 292, 1

\bibitem{} van Kerkwijk M.H., Kulkarni S.R., 2001a, A\&A, 378, 986

\bibitem{}
	van Kerkwijk M.H., Kulkarni S.R., 2001b, A\&A, 380, 221

\bibitem{} Vasisht G., Kulkarni S.R., Anderson S.B., Hamilton T.T.,
Kawai N., 1997, ApJ, 476, L43

\bibitem{} Walter F.M., Matthews L.D., 1997, Nature, 389, 358

\bibitem{} Walter F.M., Lattimer J.M, 2002, ApJ, submitted (astro-ph/0204199)

\bibitem{} Walter F.M., Wolk S.J., Neuh\"auser R., 1996, Nature, 379, 233

\bibitem{} Zampieri L., Campana S., Turolla R., Chieregato M.,
Falomo R., Fugazza D., Moretti A., Treves A., 2001, A\&A, 378, L5

\bibitem{} Zane S., Haberl F., Cropper M., Zavlin V.E., Lumb D.,
Sembay S., Motch S., 2002, MNRAS, in press

\bibitem{}
	Zavlin V.E., Pavlov G.G, 1998, A\&A, 329, 583 [ZP98]

\bibitem{}
	Zavlin V.E., Pavlov G.G., 2002, this volume [ZP02]

\bibitem{} Zavlin V.E., Pavlov G.G., Tr\"umper J., 1998, A\&A, 331, 821

\bibitem{} Zavlin V.E., Pavlov G.G., Sanwal D., Tr\"umper J., 2000,
ApJ, 540, L25

\bibitem{}
	Zavlin V.E., Pavlov G.G., Sanwal D., Tr\"umper J., Manchester R.N.,
	Halpern J.P., Becker W., 2002, ApJ, 569, 894

\end{thebibliography}
\end{document}